\documentclass[cameraready]{Interspeech}
\usepackage{amsmath}
\usepackage{amssymb}
\usepackage{algorithm}
\usepackage{algorithmic}
\usepackage{booktabs}
\usepackage{multirow}
\usepackage{graphicx}
\usepackage{cite}
\usepackage{xcolor} 
\usepackage{subcaption}
\usepackage{caption}
\usepackage{hyperref}
\usepackage{xurl}

\title{Enhancing Flow Matching with A Unified Guidance Framework for Efficient and Robust Speech Synthesis}
\author[equalcontribution]{Zuda}{Yu}
\author[equalcontribution]{Qianhui}{Xu}
\author[correspondingauthor]{Ting}{Chen}
\author[correspondingauthor]{Junhui}{Zhang}
\author[correspondingauthor]{Tao}{Fu}
\author[correspondingauthor]{Hongjiang}{Yu}
\author[correspondingauthor]{Qiangqing}{Wang}
\author[correspondingauthor]{Yang}{Song}

\address{
    Zuoyebang, China
}

\email{\{yuzuda, xuqianhui02, chenting08, zhangjunhui04, futao01, yuhongjiang, wangqiangqiang, songyang\}@zuoyebang.com}


\usepackage{comment}

\begin{document}

\maketitle

\begin{abstract}
Flow Matching (FM) has emerged as a powerful paradigm for speech generation but remains constrained by high inference latency and timbre leakage. To address these bottlenecks, we propose a unified guidance framework that enhances generation efficiency and robustness through two complementary strategies. On the data front, we introduce Data-guidance via heterogeneous augmentation, encouraging the model to disentangle linguistic content from acoustic residue. In parallel, we propose an enhanced Model-guidance mechanism that synergizes trajectory rectification with a novel intrinsic guidance objective. This approach distills conditional knowledge into network weights and straightens inference trajectory path, thereby eliminating Classifier-Free Guidance (CFG) overhead. Experiments demonstrate that our framework accelerates inference by nearly three times while effectively improving speaker similarity compared to state-of-the-art baselines. Audio samples are available here.\footnote{ \url{https://yuzuda283.github.io/unified-guidance-flow-matching/Interspeech2026_demo_samples/}}.
\end{abstract}

\section{Introduction}

Flow Matching (FM)~\cite{fm_original_paper} has rapidly advanced speech synthesis by modeling the continuous transformation from a simple prior to complex data distributions. This paradigm has demonstrated remarkable potential across various speech generation tasks. For instance, text-based approaches such as F5-TTS~\cite{f5tts} and Matcha-TTS~\cite{matchatts} achieve fully non-autoregressive speech generation by directly mapping text to acoustic features. Concurrently, discrete-token-based frameworks including the CosyVoice family~\cite{cosyvoice1,cosyvoice2,cosyvoice3,cosyvoice4}, MaskGCT~\cite{maskgct}, and FireRedTTS~\cite{fireredtts} apply FM to convert semantic representations into mel-spectrograms. Furthermore, recent audio-language models like Kimi-Audio~\cite{kimiaudio}, GLM-4-Voice~\cite{glm4}, and Step-Audio~\cite{stepaudio} increasingly rely on FM as a high-fidelity detokenizer for waveform reconstruction. Beyond speech synthesis, models such as Seed-VC~\cite{seedvc} and StableVC~\cite{stablevc} illustrate the capacity of FM to effectively disentangle speech attributes for voice conversion. Despite these impressive milestones, the widespread deployment of FM models in real-time scenarios remains fundamentally constrained by two distinct yet critical bottlenecks.

The first bottleneck, timbre leakage, compromises the robustness of zero-shot generation. Token-based models rely on discrete semantic representations as conditioning inputs, yet extracting purely disentangled content is notoriously difficult. Traditional information bottlenecks such as Vector Quantization (VQ)~\cite{vqvqe} attempt to strip away speaker identity but often degrade linguistic intelligibility and pronunciation quality. Consequently, data-driven approaches have gained prominence. Seed-TTS~\cite{seedtts} mitigates leakage by retraining models on synthetic cross-speaker pairs generated via self-distillation, while Seed-VC~\cite{seedvc} successfully employs an external generative model to perturb the source audio during training, forcing the network to align with the target prompt. These strategies inspire us to extend single-stage perturbation into a more comprehensive data-guidance constraint.

The second bottleneck, inference latency, stems from the intrinsic nature of ODE-based generation imposing two distinct computational costs. First, iterative ODE solvers require a high Number of Function Evaluations (NFE) along curved probability paths. Extensive research has focused on accelerating this process through trajectory manipulation, with fundamental techniques like Rectified Flow~\cite{reflow} attempting to straighten the probability paths. Similar concepts mapping noise to data in fewer steps are explored in InstaFlow~\cite{instaflow} and Consistency Models~\cite{consistency}. In the speech domain specifically, acceleration has been adapted through progressive distillation in ProDiff~\cite{prodiff}, adversarial training in FastDiff~\cite{fastdiff}, and consistency distillation in CoMoSpeech~\cite{comospeech}. While these trajectory-level acceleration methods substantially reduce NFE, the secondary computational burden of Classifier-Free Guidance (CFG)~\cite{cfg_paper} often remains. CFG ensures strict prompt adherence but doubles generation cost by requiring conditional and unconditional passes per sampling step. Concurrently, recent work on Model-guidance~\cite{model_guidance} offers an inspiring direction by explicitly optimizing a guided velocity field to replace CFG. Motivated by these two distinct lines of acceleration, we seek to effectively synergize them to address both trajectory curvature and guidance overhead simultaneously.

We argue that overcoming these dual challenges benefits from a unified approach: robust zero-shot generation demands severing the acoustic information shortcut during training, while efficient generation demands a straight, guidance-aware flow trajectory. To this end, we propose a unified guidance framework that systematically enhances FM through two complementary pillars:

\begin{itemize}
    \item \textbf{Data-guidance (DG):} We propose a dual-stage
    heterogeneous perturbation strategy that cascades
    model-driven cross-synthesis with signal-driven acoustic
    deformations. By deliberately degrading the acoustic
    reliability of semantic tokens during training, the model
    is forced to rely more heavily on the target prompt for
    timbre generation, thereby effectively mitigating timbre leakage.
    
    \item \textbf{Enhanced Model-guidance (MG):} We propose an advanced optimization mechanism that explicitly synergizes intrinsic guidance distillation with trajectory rectification. Our method simultaneously distills the CFG-aware velocity field into the network weights and linearizes the ODE path on the fly. This strategy circumvents the CFG dual-pass overhead and enables high-fidelity sampling in minimal steps.
\end{itemize}

Through this systematic optimization, our framework accelerates inference by nearly $3\times$ while achieving superior speaker similarity and pronunciation stability compared to state-of-the-art baselines.

\section{Methodology}

\subsection{Conditional Flow Matching}
Our method builds upon Conditional Flow Matching (CFM)~\cite{fm_original_paper}, which constructs probability paths via linear Optimal Transport. Let $x_1$ denote the target speech sample and $x_0 \sim \mathcal{N}(0, I)$ be the prior noise. The probability path $x_t$ is defined via linear interpolation:
\begin{equation}
x_t = (1-t)x_0 + t x_1, \quad t \in [0, 1].
\end{equation}
The corresponding target velocity field is given by $u_t = x_1 - x_0$. CFM trains a neural network $v_\theta(x_t, t, c)$ to approximate this velocity field by minimizing:
\begin{equation}
\label{eq:cfm_base}
\mathcal{L}_{\text{CFM}}(\theta) = \mathbb{E}_{t, x_0, x_1, c} \left[ \left\Vert v_\theta(x_t, t, c) - (x_1 - x_0) \right\Vert^2 \right].
\end{equation}
During inference, samples are generated by solving the ODE $dx/dt = v_\theta(x_t, t, c)$ from $t=0$ to $1$. To strengthen the model's adherence to the target condition $c$ (e.g., semantic tokens and speaker prompts), Classifier-Free Guidance (CFG)~\cite{cfg_paper} is widely adopted. It explicitly modifies the velocity field by extrapolating away from the unconditional prediction:
\begin{equation}
\label{eq:cfg_eq}
\tilde{v}(x_t, t, c) = (1 + w) v_\theta(x_t, t, c) - w v_\theta(x_t, t, \emptyset),
\end{equation}
where $w$ is the guidance scale. While CFG effectively improves conditional fidelity, it requires two network forward passes per integration step, doubling the computational cost and exacerbating inference latency.

\subsection{Data-guidance via heterogeneous augmentation}
Timbre leakage fundamentally stems from an information routing problem during model optimization. In conditional Flow Matching (FM) for zero-shot generation, the model is guided by two distinct prompts: semantic tokens (intended for linguistic content) and acoustic prompts (e.g., mel-spectrogram, speaker embedding). However, discrete semantic tokens are rarely pure; they inherently retain residual acoustic hints from the source audio. 

During standard reconstruction training, if the semantic tokens and the target acoustic prompt share similar acoustic properties, the network tends to take an information shortcut.  Instead of learning to properly disentangle the two inputs, the model passively extracts overlapping acoustic clues directly from the semantic tokens. Consequently, during zero-shot inference, this entangled dependency causes the residual source acoustics within the tokens to overpower the target prompt, leading to severe timbre leakage.

\begin{figure*}[t]
    \centering
    \includegraphics[width=0.75\textwidth]{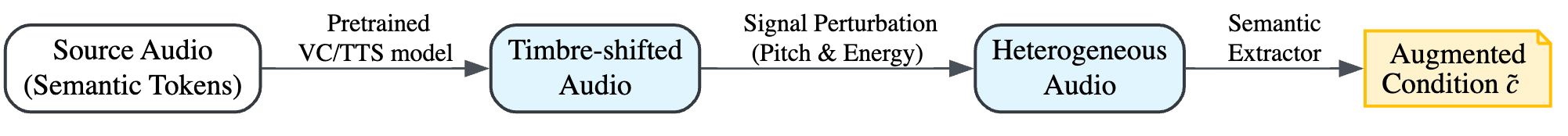} 
    \caption{Illustration of the Data-guidance (DG) strategy. DG employs a dual-stage heterogeneous perturbation pipeline. By cascading model-driven cross-synthesis with signal-driven acoustic deformation, it constructs severely mismatched training pairs. }
    \label{fig:dg}
\end{figure*}

To address this, we propose \textbf{Data-guidance (DG)}, formulated as a Dual-Stage Heterogeneous Perturbation strategy. Our objective is not merely to purify the tokens, but to deliberately degrade their acoustic reliability during training, thereby mitigating the acoustic shortcut. As illustrated in Figure~\ref{fig:dg}, we achieve this by constructing heavily mismatched training pairs through two cascaded steps:

\begin{itemize}
    \item \textbf{Model-driven cross-synthesis:} Inspired by~\cite{seedvc}, we first feed the original source semantic tokens through a generative system (e.g., a pretrained VC or TTS model) to synthesize an intermediate speech waveform. While this step introduces a preliminary identity shift, existing systems possess limited disentanglement capabilities. Consequently, the intermediate waveform inevitably retains latent acoustic residue that the model could still exploit.
    
    \item \textbf{Signal-driven acoustic deformation:} To further suppress these remaining acoustic shortcuts, we introduce a secondary perturbation. Building upon proven speech augmentation principles~\cite{dag}, we apply explicit signal-processing transformations, randomized pitch shifting and energy scaling to the intermediate waveform. This deliberate deformation effectively disrupts the residual acoustic information without altering the underlying phonetics.
\end{itemize}

The final semantic tokens, extracted from this deformed audio, serve as
the augmented condition $\tilde{c}$. During training, the FM model must
reconstruct the target speech $x_1$ from the mismatched $\tilde{c}$ and
the clean acoustic prompt. Since the acoustic properties of $\tilde{c}$
have been systematically degraded, the network is encouraged to extract
linguistic content from the semantic tokens while relying
primarily on the acoustic prompt for timbre rendering.

\subsection{Enhanced Model-guidance for unified acceleration}
The slow inference of flow matching stems from two factors: the iteration overhead due to curved ODE paths (requiring many NFE) and the guidance overhead from CFG (requiring dual forward passes). Previous works~\cite{reflow, instaflow, model_guidance} typically treat these as separate challenges.

\begin{figure*}[t]
    \centering
    \includegraphics[width=0.75\linewidth]{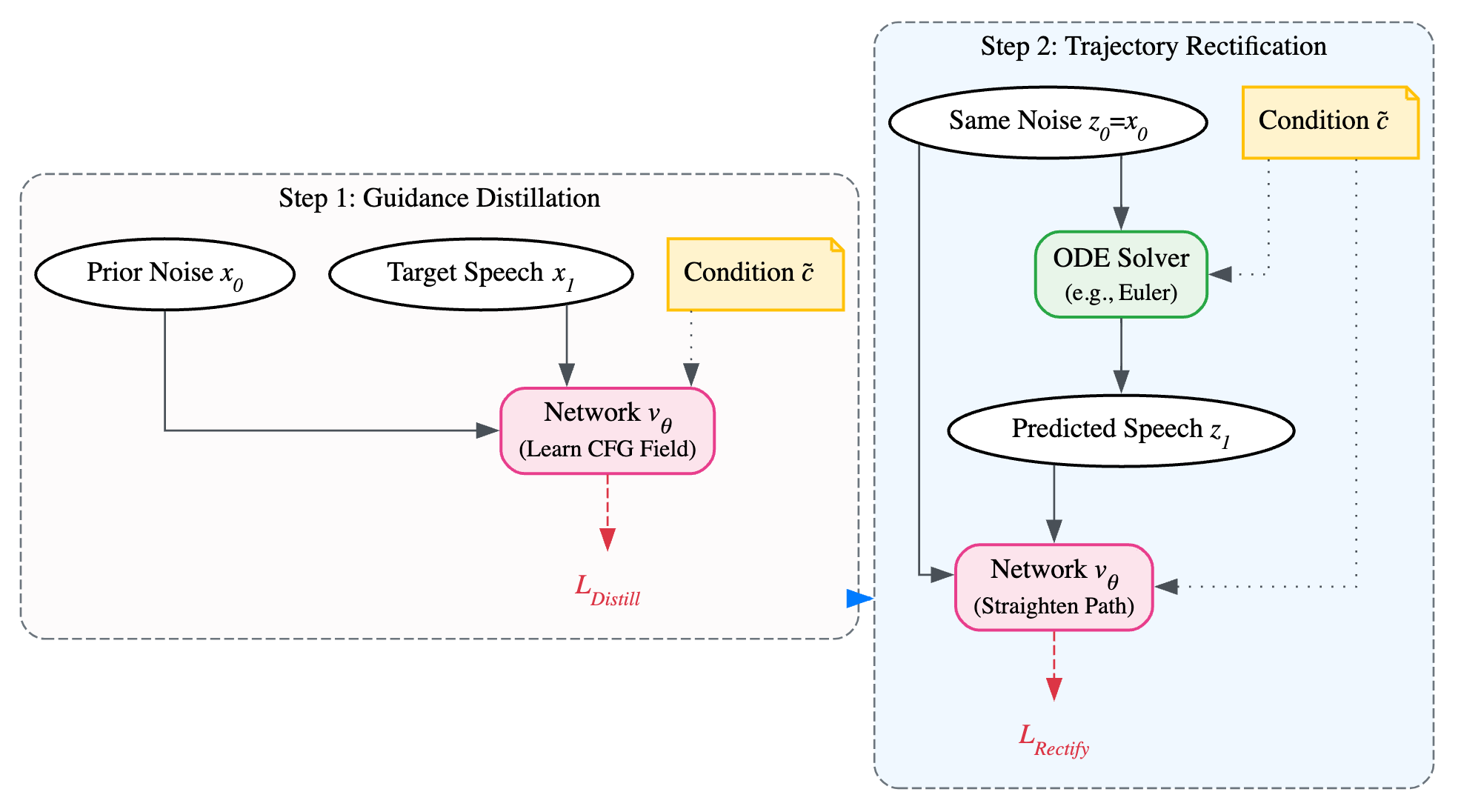} 
    \caption{Overview of the Enhanced Model-guidance (MG) mechanism. MG operates within a unified batch-level training loop. It first distills the CFG-aware vector field directly into the network weights. Subsequently, it utilizes the updated model to integrate straight trajectory paths. }
    \label{fig:mg}
\end{figure*}

We bypass this complexity with an Enhanced Model-guidance (MG) mechanism that effectively integrates intrinsic guidance distillation~\cite{model_guidance} and trajectory rectification~\cite{reflow} within a single, online training loop. As illustrated in Figure~\ref{fig:mg}, for each training batch, the optimization alternates between two sequential steps utilizing the augmented condition $\tilde{c}$.


\subsubsection{Model-guidance distillation} 
First, we infuse the CFG mechanism directly into the network weights. We formulate an intrinsic guidance objective that replaces the standard CFM target with a guided velocity field:
\begin{equation}
\label{eq:mg_target}
v'_{\text{target}} = (x_1 - x_0) + w \cdot \text{sg}\big( v_{\theta}(x_t, t, \tilde{c}) - v_{\theta}(x_t, t, \emptyset) \big),
\end{equation}
where $\text{sg}(\cdot)$ denotes the stop-gradient operation. The network is updated via a single backward pass to minimize the distillation loss:
\begin{equation}
\label{eq:mg_loss}
\mathcal{L}_{\text{Distill}}(\theta) = \mathbb{E}_{t, x_0, x_1, \tilde{c}} \left[ \left\Vert v_{\theta}(x_t, t, \tilde{c}) - v'_{\text{target}} \right\Vert^2 \right].
\end{equation}
By optimizing $\mathcal{L}_{\text{Distill}}$, the network intrinsically learns the CFG-aware vector field, enabling it to generate highly condition-aligned results using only a single forward pass $v_\theta(x_t, t, \tilde{c})$ at inference time.

\subsubsection{Trajectory rectification} 
Following the distillation step, we leverage the updated, CFG-distilled network to straighten the generation trajectory. Using the same batch of prior noise $z_0 = x_0$ and the augmented condition $\tilde{c}$, we perform numerical ODE integration to predict the corresponding clean speech feature $\hat{z}_1$:
\begin{equation}
\hat{z}_1 = \operatorname{ODE}(v_{\theta}, z_0, \tilde{c}).
\end{equation}
Because the current model $v_\theta$ has already absorbed the CFG knowledge, this generation does not require double forward passes, making the on-the-fly simulation efficient. We then construct a rectified, straight probability path $z_t = (1-t)z_0 + t \hat{z}_1$ and update the network to track this linear trajectory:
\begin{equation}
\label{eq:joint_loss}
\mathcal{L}_{\text{Rectify}}(\theta) = \mathbb{E}_{t, z_0, \hat{z}_1, \tilde{c}} \left[ \left\Vert v_{\theta}(z_t, t, \tilde{c}) - (\hat{z}_1 - z_0) \right\Vert^2 \right].
\end{equation}

\section{Experiments}
\subsection{Experimental setup}

\subsubsection{Datasets}
We employ a two-stage data construction strategy to balance pre-training scale with optimization quality. First, to establish a robust initialization, the foundation model is pre-trained on 50k hours of English speech from the Emilia dataset~\cite{emilia} under a standard matched-condition setting. Second, to enforce the proposed Data-guidance (DG) constraints, we curate a high-quality subset of 30k hours by filtering for DNSMOS~\cite{dnsmos} scores above 3.2. We then apply our dual-stage heterogeneous perturbation pipeline (model-driven cross-synthesis followed by randomized pitch and energy shifting) to this subset. This generates an additional 30k hours of acoustically mismatched but linguistically identical pairs. The original and perturbed subsets are combined to form a 60k-hour mixed corpus, which serves as the training data for all subsequent Enhanced Model-guidance (MG) optimization stages.

\subsubsection{Experimental configuration}
\noindent\textbf{Model architecture:}
We build a token-based flow matching model inspired by
CosyVoice2~\cite{cosyvoice2}. Unlike its original
convolution-augmented architecture, our decoder adopts a pure
Diffusion Transformer (DiT)~\cite{dit}. The model consists of
20 stacked DiT blocks, each configured with an attention
dimension of 1024 and a feed-forward network (FFN) dimension
of 4096, totaling approximately 330M parameters. Instead of
concatenating speaker embeddings with the input tokens, we
inject speaker information through Adaptive Layer
Normalization (AdaLN)~\cite{adaln}, which provides more
effective and fine-grained timbre control in practice.

\noindent\textbf{Training details:}
All training processes are conducted on 16 NVIDIA H100 GPUs.
Our foundation Flow Matching model is first pre-trained
for 5 epochs. We optimize the network using AdamW, with
the learning rate linearly warmed up to a peak of
$1 \times 10^{-4}$ over the first 1,000 steps, followed
by a cosine annealing decay to $1 \times 10^{-5}$.
This pre-training stage takes approximately 48 hours.

Building upon this initialization, we deploy our unified optimization pipeline on the 60k-hour mixed corpus. The Enhanced MG is executed for 2 epochs. Specifically, to achieve true joint optimization, we perform two distinct backward passes for each mini-batch: First, we compute and backpropagate the Intrinsic Guidance Distillation loss (Equation~\ref{eq:mg_loss}) to internalize the CFG field. Second, we immediately utilize the updated weights to simulate straight trajectories on the fly with the exact same batch of noise, backpropagating the Rectification loss (Equation~\ref{eq:joint_loss}). During this unified optimization phase, we freeze all modules except the DiT decoder to ensure the model efficiently learns the straightened flow paths without catastrophically forgetting its pre-trained linguistic representations. Due to the online training, this stage takes approximately 90 hours.

\subsection{Evaluation metrics}
To assess the performance of our guidance-enhanced Flow Matching, we consider the following objective metrics:

\noindent\textbf{Real-time factor (RTF):} Measures inference speed. We report RTF for the Flow Matching model alone, isolating its impact from other pipeline components.

\noindent\textbf{Speaker similarity (SIM):} Cosine similarity between speaker embeddings of generated speech and reference prompts. We extract embeddings using Cam++ model~\cite{camplus}.

\noindent\textbf{Word error rate (WER):} Assesses linguistic intelligibility by comparing Whisper-Large ASR~\cite{whisper} transcriptions with the ground-truth text.

\subsection{Results and discussion}
We report evaluation results separately for Voice Conversion (VC) and Text-to-Speech (TTS) to highlight task-specific performance. All samples are generated on an NVIDIA RTX 4090 GPU. VC is conditioned on semantic tokens from source speech, while TTS uses semantic features derived from text via the CosyVoice2 LLM \cite{cosyvoice2} and synthesizes audio with the HiFTNet Vocoder~\cite{hifinet}.

\subsubsection{Voice Conversion (VC) performance}

Table~\ref{tab:vc_results} presents the ablation results for Voice Conversion under both \textbf{Parallel} (same speaker, different audio) and \textbf{Non-Parallel} (cross-speaker) settings. We first isolate the effect of Data-guidance (DG). Applying DG alone yields the highest Speaker Similarity (SIM) scores, exceeding the Ground Truth SIM, suggesting that our heterogeneous perturbation effectively encourages the model to ignore source acoustic residue. This confirms that our heterogeneous perturbation successfully forces the model to ignore source acoustic residue. Conversely, addressing only the inference bottleneck via Model-guidance (MG) reduces latency but incurs trade-offs. While Vanilla MG internalizes the CFG logic to lower the RTF from 0.078 to 0.058, our Enhanced MG further incorporates trajectory rectification to achieve an extreme 3-step inference. However, without data-level correction, this acceleration leads to a minor degradation in SIM.

Finally, our unified guidance framework resolves this dilemma by combining the 3-step Enhanced MG with the DG strategy. Although its absolute SIM is marginally lower than the computationally heavy DG-only model, it achieves a substantial $3.25\times$ speedup, mitigating the performance drop seen in the pure Enhanced MG model and outperforming the 10-step Base baseline. Crucially, the Unified framework demonstrates strong zero-shot robustness: its Non-Parallel SIM on LibriTTS (0.808) not only surpasses the Base model but even exceeds the Ground Truth Parallel SIM (0.799). This confirms that despite operating under 3-step acceleration without CFG, our framework achieves highly effective content-timbre disentanglement, offering a practical solution for real-time deployment.


\begin{table}[htbp]
\centering
\caption{Results for Voice Conversion on LibriTTS and Seed-TTS. The table demonstrates the individual and combined impact of the Data-guidance (DG) strategy and Enhanced Model-guidance (MG) mechanism. }
\label{tab:vc_results}
\resizebox{\columnwidth}{!}{
\begin{tabular}{l|c|cc|cc}
\toprule
\multirow{3}{*}{\textbf{Method}} & \multirow{3}{*}{\textbf{RTF ($\downarrow$)}} & \multicolumn{4}{c}{\textbf{SIM ($\uparrow$)}} \\
\cmidrule{3-6}
 & & \multicolumn{2}{c|}{\textbf{Parallel}} & \multicolumn{2}{c}{\textbf{Non-Parallel}} \\
\cmidrule{3-6}
 & & \textbf{LibriTTS} & \textbf{Seed-TTS} & \textbf{LibriTTS} & \textbf{Seed-TTS} \\
\midrule
Reference (GT) & - & 0.799  & 0.789 & 0.073 & 0.128 \\
\midrule
Base Model (10 NFE) & 0.078 & 0.874 & 0.800 & 0.793 & 0.730 \\
\midrule
\multicolumn{6}{l}{\textit{Data-guidance (DG)}} \\
\quad+ DG (10 NFE) & 0.078 & \textbf{0.897}  & \textbf{0.822} & \textbf{0.869} & \textbf{0.792} \\
\midrule
\multicolumn{6}{l}{\textit{Model-guidance (MG)}} \\
\quad+ Vanilla MG (10 NFE) & 0.058 & 0.885  & \underline{0.810} & 0.813 & 0.744 \\
\quad+ Enhanced MG (3 NFE) & \textbf{0.024} & 0.870  & 0.791 & 0.792 & 0.722 \\
\midrule
\textbf{Unified Guidance (3 NFE)} & \textbf{0.024} & \underline{0.887} & 0.808 & \underline{0.850} & \underline{0.767} \\
\bottomrule
\end{tabular}
}
\end{table}

\begin{table}[th]
\centering
\caption{Text-to-Speech (TTS) performance on LibriTTS and Seed-TTS.}
\label{tab:tts_results}
\resizebox{\columnwidth}{!}{
\begin{tabular}{l|cc|cc}
\toprule
\textbf{Method} & \multicolumn{2}{c|}{\textbf{LibriTTS}} & \multicolumn{2}{c}{\textbf{Seed-TTS}} \\
 & \textbf{WER ($\downarrow$)} & \textbf{SIM ($\uparrow$)} & \textbf{WER ($\downarrow$)} & \textbf{SIM ($\uparrow$)} \\
\midrule
Reference (GT) & 2.12 & 0.799 & 1.82 & 0.789 \\
\midrule
CosyVoice2 & \textbf{2.57} & 0.847 & 2.47 & 0.750 \\
Base Model & \textbf{2.57} & 0.871 & \textbf{2.22} & 0.794 \\
\textbf{Unified Guidance} & 2.60 & \textbf{0.888} & 2.45 & \textbf{0.806} \\
\bottomrule
\end{tabular}
}
\end{table}

\subsubsection{Text-to-Speech (TTS) performance}
To evaluate the generalization capability of our framework, we extend our experiments to the zero-shot TTS task. We utilize the CosyVoice2 LLM \cite{cosyvoice2} to generate semantic tokens and exclusively swap the Flow Matching (FM) model. 

As reported in Table~\ref{tab:tts_results}, our 3-step Unified FM backend maintains comparable intelligibility compared to the official CosyVoice2 backend. While extreme trajectory rectification introduces a slight degradation in WER, the linguistic stability remains robust. More importantly, when paired with the same LLM, our unified model achieves a higher SIM score outperforming the unoptimized Base FM model. This demonstrates that our Data-guidance strategy effectively filters out latent acoustic variance inherently predicted by the LLM, anchoring the timbre to the target prompt. Overall, our framework serves as a robust, efficient acoustic detokenizer for existing TTS systems.

\section{Conclusion}
This paper presented a unified guidance framework that addresses the distinct bottlenecks of timbre
leakage and high inference latency in Flow Matching-based
speech generation. On the data side, our Data-guidance
strategy employs dual-stage heterogeneous perturbation to
sever acoustic shortcuts during training, promoting robust
content-timbre disentanglement. On the model side, our
Enhanced Model-guidance mechanism integrates intrinsic
guidance distillation with online trajectory rectification,
entirely eliminating CFG overhead. Experiments on both Voice Conversion and
Text-to-Speech benchmarks show that the proposed framework
achieves a $3.25\times$ inference speedup while consistently improving zero-shot speaker similarity.
Furthermore, when deployed as an acoustic backend for
existing TTS systems, our framework maintains competitive
intelligibility while delivering improved speaker fidelity.
These results demonstrate that our unified approach offers
a practical and effective solution for real-time speech
generation. 
\section{Acknowledgments}
\ifcameraready
    The authors thank the Zuoyebang Speech Team for providing the computing power and related platforms that made this research possible. We also acknowledge the open-source community for the models and datasets used in this study.
\else
    (Hidden for review)
\fi

\section{Generative AI Use Disclosure}
The authors used generative AI tools (Gemini 3.0 pro) solely for the purpose of checking LaTeX formatting, correcting syntax errors, and refining the layout of this manuscript. No AI tools were used to generate scientific content, experimental data, or the intellectual ideas presented in this work.

\bibliographystyle{IEEEtran}
\bibliography{mybib}

\end{document}